\documentclass[twocolumn,showpacs,preprintnumbers,amsmath,amssymb]{revtex4}

\usepackage{graphicx}
\usepackage{dcolumn}
\usepackage{bm}

\begin{document}
\newcommand{\braopket}[3]{\langle #1 | \hat #2 |#3\rangle}
\newcommand{\braket}[2]{\langle #1|#2\rangle}
\newcommand{\bra}[1]{\langle #1|}
\newcommand{\braketbraket}[4]{\langle #1|#2\rangle\langle #3|#4\rangle}
\newcommand{\braop}[2]{\langle #1| \hat #2}
\newcommand{\ket}[1]{|#1 \rangle}
\newcommand{\ketbra}[2]{|#1\rangle \langle #2|}
\newcommand{\op}[1]{\hat {#1}}
\newcommand{\opket}[2]{\hat #1 | #2 \rangle}
\preprint{APS/123-QED}

\title{Carbon nanotube:  a low-loss spin-current waveguide. }

\author{F. S. M. Guimar\~aes$^a$, D. F. Kirwan$^b$, A. T. Costa$^{a}$, R. B. Muniz$^{a}$, D. L. Mills$^c$ and M. S. Ferreira$^{b}$
}
\affiliation{
(a) Instituto de F\'{\i}sica, Universidade Federal Fluminense, Niter\'oi, Brazil \\
(b) School of Physics, Trinity College Dublin, Dublin 2, Ireland \\
(c) Department of Physics and Astronomy, University of California, Irvine, California 92697, USA}

\date{\today}

\begin{abstract}
We demonstrate with a quantum-mechanical approach that carbon nanotubes are excellent spin-current waveguides and are able to carry information stored in a precessing magnetic moment for long distances with very little dispersion and with tunable degrees of attenuation. Pulsed magnetic excitations are predicted to travel with the nanotube Fermi velocity and are able to induce similar excitations in remote locations. Such an efficient way of transporting magnetic information suggests that nanotubes are promising candidates for memory devices with fast magnetization switchings. 

\end{abstract}

\maketitle
\bibliographystyle{apsrev} 

The issue of magnetization dynamics in low-dimensional systems is currently one of the most studied in spintronics \cite{book}. In particular, controlling how quickly the energy of a precessing magnetic moment propagates through wires is motivated by the possibility of building miniaturized memory devices with fast magnetization switchings. This requires structures functioning as efficient spin-current waveguides, {\it i.e.}, conduits capable of transporting spin information with little attenuation. Renowned for their high thermal and electronic conductivities, for their long aspect ratios and coherence lengths, carbon nanotubes (NT) possess the ideal characteristics for acting as waveguides. NT have indeed been shown to function as conduits for electrons \cite{wave1, wave2} and for phonons \cite{wave3}. A natural question to ask, raised in this manuscript, is whether these materials are also efficient spin-current waveguides. 

Tserkovnyak \cite{Tser, bauer-e-cia} proposed a mechanism for pumping spin current into a nonmagnetic metal caused by the precession of an adjacent magnetization. In this case, angular momentum from the moving magnetization is transferred to the conduction electrons, creating a spin disturbance that propagates throughout the metallic conduit. This spin flow is produced without an applied voltage, involves no net electrical current, and may be used to excite other magnetic units also in contact with the non-magnetic metal. Our strategy is to test this mechanism with NT in contact to local magnetic moments. 

Obstacles for identifying good spin waveguides lie in the difficulties of predicting the precise time evolution of the magnetization of a realistic system, taking into consideration the details of its electronic structure. This requires a full understanding of how a pulsed magnetic excitation evolves. Although previously treated semiclassically \cite{Tser, bauer-e-cia}  and quantum-mechanically \cite{njp,prb-08}, a time-dependent quantum approach for this problem is still missing. In this manuscript we bridge this gap and provide a quantum formalism that describes in time domain how a localized magnetic perturbation propagates throughout a nanoscale system, more precisely a metallic NT. In doing so, we are able to show that NT are excellent spin-current waveguides. We demonstrate that they transport spin current across long distances in a dispersionless fashion and with little attenuation. Moreover, we indicate that pulsed spin excitations propagate with a characteristic speed given by the electronic Fermi velocity and with a tunable degree of attenuation that depends on the excitation frequency. 

\begin{figure}
\includegraphics[width = 8.6cm,clip]{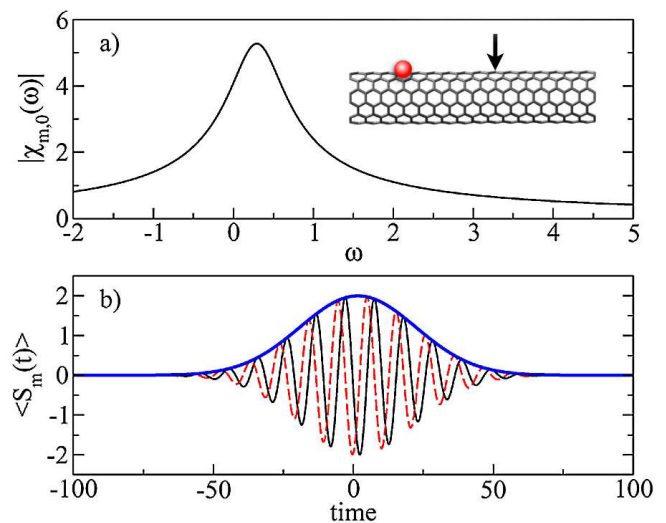}
\caption{(a) Spin susceptibility $\chi_{m,0}$ (in units of $\hbar/{\rm eV}$) as a function of the excitation frequency $\omega$ (in ${\rm THz}$) for a (4,4) NT. The distance between site $m$ and the impurity is $x_m=3a_0$. Inset shows NT with a single magnetic impurity. Arrow represents where the electrons spin is probed. (b) Solid and dashed lines represent the $\hat{x}$- and $\hat{y}$-components of the spin disturbance $S^+_m(t)$; the thick solid line depicts $|S^+_m(t)|$. Spins are in units of $\hbar g \mu_B h_0$, where
$g$ is the electron spin g-factor, $\mu_B$ is the Bohr magneton and $h_0$ is the
amplitude of the applied time-dependent transverse field; time is given in ps.
Results were generated with a gaussian pulse centered at frequency $\omega = 0.6
{\rm THz}$ and with a standard deviation $\sigma = 20.7 \, {\rm ps}$.  }
\label{figure_1}
\end{figure}

We consider a metallic NT doped with a single substitutional magnetic impurity \cite{sanchez-portal, prl-arkady}, as depicted in the inset of figure 1. Assuming the existence of a localized magnetic moment pointing along the
$\hat{z}$-direction, we consider a perturbing time-dependent transverse magnetic field ${\vec h}_\perp(t)$ which sets the impurity magnetic moment into precession. Our treatment applies to magnetic moments at temperatures above their Kondo temperatures which are nevertheless expected to be quite low in NT. We assume the Fermi-liquid representation and do not include any Luttinger-liquid effects. To determine whether a NT is a good conduit for spin currents, we must assess how such a localized magnetic excitation propagates across the structure. In other words, we must investigate how this excitation disturbs the spin balance of the system not only where the impurity is located but also, and more importantly, how the local spin dynamics is affected away from the impurity. This is naturally manifested by the spin susceptibility $\chi$, which reflects how the spin degrees of freedom of a system responds to a magnetic excitation. 

To calculate the spin susceptibility one needs the Hamiltonian describing the electronic structure of the unperturbed system, which we assume is given by  $\hat{H} =   \sum_{<i,j>,\sigma} \gamma_{ij} \, \ {\hat
c}_{i\sigma}^\dag \, {\hat c}_{j\sigma} + \sum_\sigma (\epsilon_0 \, {\hat
n}_{0,\sigma} + {U \over 2}  \, {\hat n}_{0\sigma} \, {\hat n}_{0{\bar \sigma}})
+ \hat{H}_Z$. Here, $\gamma_{ij} $ represents the electron hopping between
nearest neighbor sites $i$ and $j$,  $\hat{c}_{i\sigma}^{\dag}$ creates an
electron with spin $\sigma$ in site $i$, $\epsilon_0$ is the atomic energy level
of the magnetic impurity occupying the site $i=0$, $\hat{n}_{0 \sigma} =
\hat{c}_{0\sigma}^{\dag} \hat{c}_{0 \sigma}$ is the corresponding electronic
occupation number operator, and $U$ represents an effective on-site interaction
between electrons in the magnetic site, which is neglected elsewhere. Finally,
$H_Z$ plays the role of a local Zeeman interaction that defines the
$\hat{z}$-axis as the equilibrium direction of the magnetization. The
Hamiltonian parameters can be obtained from density-functional-theory
calculations so that the electronic structure of the doped system
is well described \cite{charlier-prl04, rocha}. Although we shall present
results for a NT doped with Mn atoms, other substitutional magnetic impurities
may be employed \cite{prl-arkady}. In our calculations we fix the Fermi energy
$E_{F}=0$, and use $\gamma_{C,C} = 2.7$ eV, and $\gamma_{Mn,C} = 1.0$ eV. We
take the number of $d$-electrons in the Mn site $n_0 = 1$, $\epsilon_0=0$, $U=5$
eV, and assume a local Zeeman energy splitting $\delta_0=1$ meV.  Spin-orbit coupling is neglected for being very small compared to the electronic bandwidth  of NT. 

The time-dependent transverse spin susceptibility is defined as $\chi_{m,j}(t) =
-{i \over \hbar} \Theta(t)\langle[{\hat S}_m^+(t),{\hat S}_j^-(0)]\rangle$,
where $\Theta(x)$ is the heaviside step function, and ${\hat S}_m^+$ and ${\hat
S}_m^-$ are the spin raising and lowering operators at site $m$, respectively.
The indices $j$ and $m$ refer to the locations where the field is applied and
where the response is measured, respectively. In our case, a precession of the
magnetic moment is induced at site $j=0$, and we wish to observe the spin
disturbance at an arbitrary site $m$. This response is fully described by
$\chi_{m,0}(t)$. Within the random phase approximation, this susceptibility may
be calculated in frequency domain, and in matrix form it is given by
$\chi(\omega) = [1 + \chi^0(\omega)\, U]^{-1} \, \chi^0(\omega)$, 
where $\chi^0$ is the Hartree-Fock susceptibility, whose
matrix elements are given by \begin{widetext}
\begin{equation}
\chi_{m,j}^0(\omega) = {i \hbar \over 2 \pi} \int_{-\infty}^{+\infty} d\omega^\prime f(\omega^\prime) \left\{ \left[g_{j,m}^\uparrow(\omega^\prime) -  g_{m,j}^{\uparrow *}(\omega^\prime)  \right]  g_{m,j}^\downarrow(\omega^\prime+\omega) +  \left[g_{m,j}^\downarrow(\omega^\prime) -  g_{j,m}^{\downarrow *}(\omega^\prime)  \right]  g_{m,j}^{\uparrow *}(\omega^\prime+\omega) \right\}\,\,.
\label{susc-hf}
\end{equation}
\end{widetext}
Here, $g_{m,j}^{\sigma}(\omega)$ represents the time Fourier transforms of the
retarded single-particle propagators for an electron with spin $\sigma$ between
sites $m$ and $j$, and $f(\omega)$ is the Fermi function. For
a pristine NT $g_{mj}^{\sigma}(\omega)$ may be analytically determined
\cite{coupling1}, and the integration in Eq. (\ref{susc-hf}) performed with
great numerical accuracy. 

It is instructive to start by calculating $\chi_{m,0}(\omega)$ given by Eq. (\ref{susc-hf}). A typical result is shown in Fig. 1a for a (4,4) armchair NT doped with a single Mn impurity. It depicts
$|\chi_{m,0}(\omega)|$ which is proportional to the amplitude of the spin
disturbance at site $m$ due to a time-dependent transverse magnetic field
applied at site $0$. In Fig. 1a the site $m$ is a distance $x_m=3a_0$ from the
impurity, where $a_0\approx 2.46$\AA\, represents the graphene lattice
parameter. A very distinctive peak is evident in the $\omega$-dependent
susceptibility, reflecting the existence of a resonance close to the Larmour
frequency, as expected. 

\begin{figure}
\includegraphics[width = 8.6cm,clip]{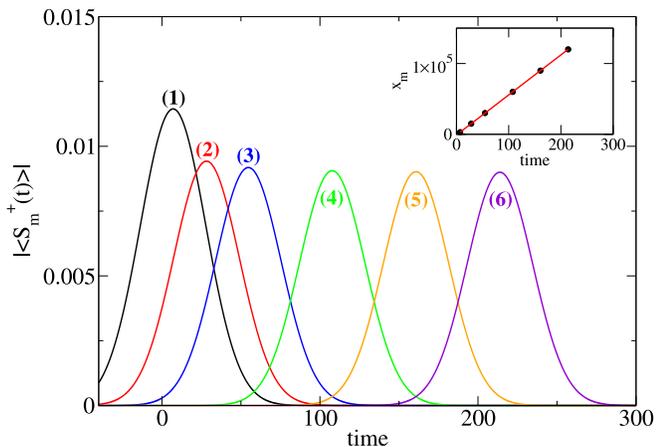}
\caption{Amplitude of the transverse spin disturbance (in units of $\hbar g \mu_B h_0$) as a function
of time (in ps) at different distances $x_m$ from the magnetic impurity. The curves
labelled from (1) to (6) represent signals probed at different values of $x_m
=0.3, 1.5, 3, 6, 9,$ and $12 \times 10^4 a_0$, respectively. Inset plots how
the corresponding probing positions $x_m$ relate with the time $\tau$ taken to
reach a maximum spin disturbance.} \label{figure_2}
\end{figure}

Since we are interested in studying the propagation of magnetic signals, it is illustrative to look at the susceptibility as a function of time rather
than frequency. The Fourier transform of the curve depicted in Fig. 1a
gives the spin disturbance response to an instantaneous $\delta$-like pulse
$\vec{h}_{\perp}(t)$ applied at site $0$. However, for practical reasons, we
consider the response to a gaussian transverse magnetic field pulse given by
${\vec h}_\perp(t) = {\vec h}_0 \, \exp(-{t^2 \over 2 \sigma^2})$ applied along
the $\hat{x}$-direction ($\vec{h}_0=h_0\hat{x}$). Here $h_0$ represents the
maximum field strength and  $\sigma$ its standard deviation. Fig. 1b shows how such a magnetic pulse applied at the impurity location affects the spin balance of the NT at a nearby site
($x_m=3a_0$). The thin solid and dashed lines represent the calculated spin
components $\langle S_m^x(t) \rangle$ and $\langle S_m^y(t) \rangle$,
respectively, and the modulating thick solid line is the magnitude of the
transverse spin disturbance given by $\vert \langle S_m(t)\rangle \vert \equiv
\sqrt{\langle S_m^x \rangle^2 + \langle S_m^y \rangle^2}$.  To avoid congested
figures the magnitude will be hereafter used to represent the spin
dynamics on the NT. 

The fact that the electron spin probed on the NT displays the precessional
motion originated at the impurity is a clear indication of a flowing
spin current emanating from the impurity into the NT. Earlier calculations have suggested that the existence of this current is sufficient to induce a dynamic magnetic coupling between
dispersed impurities in NT, which is far more pronounced and
long-ranged than any other magnetic coupling of static nature \cite{coupling1,
carbon,njp,prb-08}. To test how this precession propagates, in Figure 2
we show $\vert \langle S_m(t)\rangle \vert$ probed at
different locations on the NT. The six numbered lines correspond to
spin disturbances probed at different sites along a line parallel to
the NT axis containing the impurity site $0$. The spin disturbance probed on
other sites belonging to the same NT ring are virtually identical to the ones
displayed in Fig. 2. Small deviations are noticeable when the ring is
very close to the impurity but disappear after a short propagation
distance. This indicates that the initial omni-directional excitation produced
by the applied perturbation decays very fast into a cylindrical wave front that moves
along the NT axial direction with a uniform speed. This propagation speed is found from the slope of the straight line in the inset of Fig. 2 showing the probing position $x_m$ as a function of the time $\tau$ spent by the pulse maximum to reach $x_m$. We find the propagation velocity $v =  1.4 \times 10^5 \, {\rm m/s}$, which is precisely the NT Fermi velocity, indicating that  the conduction electrons at $E_F$ are the main carriers of the spin current. 

Figure 2 shows that, as the pulse moves, it is initially deformed but it quickly reaches a form that remains unaltered for asymptotically long distances. This can be seen when we focus on two features of the pulsed signals, namely their maximum amplitude $\vert \langle S_m\rangle \vert_{\rm max}$ and their corresponding width, here represented by the mid-height separation $\Delta$. Figure 3a plots $\vert \langle S_m\rangle
\vert_{\rm max}$ and $\Delta$ as a function of the probing position $x_m$ in the
case of an excitation pulse whose frequency spectrum is centred at the same
frequency as used for obtaining Fig. \ref{figure_1}b. Both quantities saturate
after a short distance and are stationary for $x_m > 10^4 \, a_0$. This is a
remarkable result showing that the information contained in the magnetic
pulse can be transported across very long distances without distortion.
Surely, the asymptotic value for $\vert \langle S_m\rangle \vert_{\rm max}$ is
not the same as the one probed near the impurity but it still has a sizable
magnitude indicating that a considerable fraction of the energy contained in the
magnetization precession can be used elsewhere. 

\begin{figure}
\includegraphics[width = 8.6cm,clip]{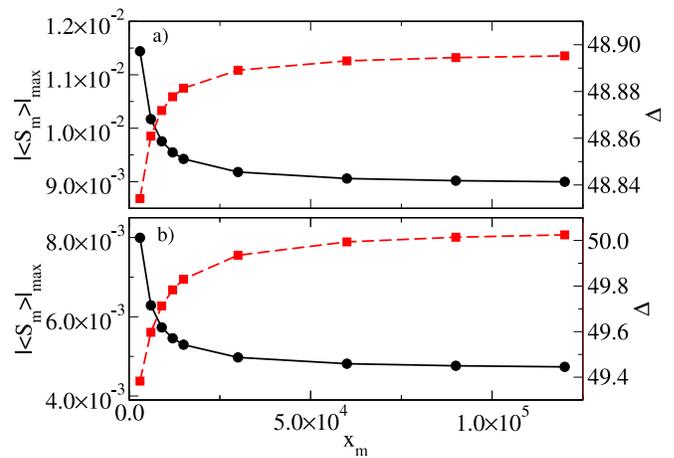}
\caption{Left axis depicts maximum amplitude of the pulsed excitation
$\vert \langle S_m\rangle \vert_{\rm max}$ (in units of $\hbar g \mu_B h_0$) probed at site $m$ and is shown by
the solid line with circular symbols. Right axis represents the width $\Delta$ (in ps)
of the probed pulses shown by dashed lines with square symbols.
Both quantities are plotted as a function of the separation $x_m$ (in units of
$a_0$) from the impurity location. Panels (a) and (b) are for pulses centered at
resonance and off-resonance frequencies, respectively. } \label{figure_3}
\end{figure}

Furthermore, we can also control the fraction of the original pulse that reaches this asymptotic regime by selecting the central frequency in the excitation pulse. Figure 3b displays $\vert \langle S_m\rangle \vert_{\rm max}$ and $\Delta$ for the case of a pulse whose spectrum is centered at an
off-resonance frequency $\omega = 0.24 \, {\rm THz}$. In this case the
asymptotic value for $\vert \langle S_m\rangle \vert_{\rm max}$ is much
reduced when compared with its counterpart in Fig. 3a. The existence of this
non-decaying asymptotic contribution to the spin-current has been demonstrated
recently for NT and atomic chains \cite{njp, prb-08} but here we explicitly show
that this can be actually controlled by an appropriate selection of excitation
frequencies. All the features presented so far indicate that NT can carry
spin-current information contained within pulsed excitations without dispersing
their form. This can be explained by the peculiar electronic structure exhibited
by NT, which is notoriously dispersionless at the Fermi level. In other words,
extended states at the Fermi level have a linear dependence on the electronic
wave vectors. In this case, group and phase velocities are interchangeable,
meaning that every single frequency-component comprising  a pulse will travel
with the same speed, thus preventing any distortion of the pulse shape. If NT
are to be used as spin-current waveguides, these features must be tested in the
presence of structural disorder. In fact, we have simulated the presence of isolated Boron substitutional impurities to explore their effect in the propagation of the spin disturbance pulse. We find that the pulse preserves its shape and amplitude, except very close to the B impurity, where it becomes slightly distorted. We notice that at larger distances, however, the pulse shape is restored as if no impurity is present. This suggests that the spin-current waveguide features reported so far are robust against the presence of non-magnetic disorder in NT. 

\begin{figure}
\includegraphics[width = 8.6cm,clip]{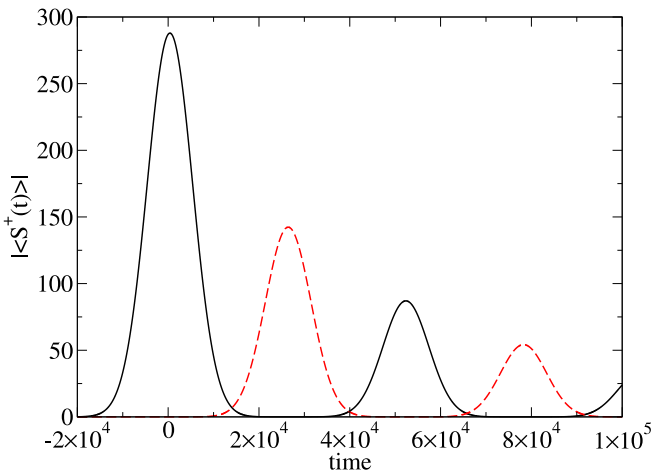}
\caption{Amplitude of the pulsed excitation as a function of time for the case
of two magnetic impurities a long distance apart ($x_m =  6 \times 10^4 a_0$).
Solid line shows the spin disturbance evaluated at the impurity where the
excitation is induced whereas the dashed line is evaluated at the other impurity
location. } \label{figure_4}
\end{figure}

Finally, finding small values for the pulse amplitude at asymptotically large
distances (see left axis of Figs. \ref{figure_3}a an \ref{figure_3}b) does not
imply that the magnetic information is lost. To make this point more explicitly
we add a second magnetic impurity a long distance from the original one ($x_m =
6 \times 10^4 a_0$) and see whether the excitation produced at the origin can be
transported to the new location. Figure \ref{figure_4} shows the spin
disturbance $\vert \langle S^+_m\rangle \vert$ probed on both impurities. The
solid line shows the magnetization precession at the origin whereas the dashed
line represents the spin disturbance as a function of time measured at the
second impurity. Both lines have equally spaced peaks that are a time interval
$\Delta t = 2 x_m/v$ apart but are shifted by half that amount. This is easily
explained by the fact that precessing magnetic moments will always emit spin
current to the surrounding conduction electrons which will induce further
precessions when interacting with other magnetic moments. Therefore, the induced
precession probed at the origin appears as the first peak in Fig. \ref{figure_4}. The subsequent pulse is seen to occur after the spin current has traveled all the way to excite the second impurity at site $m$. Once excited,
this impurity precesses also emiting spin current, which will again induce another precession at the origin. This process is repeated indefinitely each time with smaller amplitudes. The amplitude reduction is inevitable because half of the energy stored in the pulsed precession will travel away from the other
impurity. But the fact that the pulse produced at the origin reappears a long
distance apart means that the energy stored in the precession of magnetic
moments can be noiselessly transported across the conduction electrons of NT,
which confirms that these materials are ideal spin-current waveguides. 

In summary, we have shown that metallic NT are excellent spin-current
waveguides and are able to transport magnetic information across long distances
with minimum dispersion and with very little loss. Spin disturbances induced by
localized magnetic excitations are shown to propagate throughout the length of a
metallic NT with speeds of the order of $10^5$ m/s. Moreover, we show that the
energy stored in a magnetic precession can be used elsewhere when the spin
current traveling through NT interacts with other magnetic objects. These features turn NT into ideal components for fast-response memory devices. The experimental testing of our predictions requires the generation and detection of spin currents. One possibility is to excite a spin with a laser pulse, and monitor a second spin located somewhere downstream with an STM tip. Since the STM current can be modulated by a coherent spin precession, one should be able to observe the second spin being excited by the spin current in the NT.

\end{document}